\documentclass[12pt,a4paper]{article}

\usepackage[truedimen,margin=30mm]{geometry} 

\usepackage{mathrsfs}
\usepackage{amssymb}
\usepackage{amsmath}
\usepackage{ascmac}
\usepackage{amsthm}
\usepackage[dvipdfmx]{graphicx}
\usepackage{natbib}
\usepackage{setspace}
\usepackage{url}

\usepackage{color}
\usepackage{times}

\usepackage{titlesec}
\titleformat*{\section}{\large\bfseries}
\titleformat*{\subsection}{\it}

\newtheorem{thm}{Theorem}

\title{{\bf Prior Sensitivity Analysis without Model Re-fit}}

\date{}

\begin{document}

\maketitle
\doublespacing

\vspace{-1.7cm}
\begin{center}
{\large Shonosuke Sugasawa}

\medskip

\noindent
{\it Faculty of Economics, Keio University}\\
\end{center}

\vspace{0.5cm}
\begin{center}
{\bf \large Abstract}
\end{center}

Prior sensitivity analysis is a fundamental method to check the effects of prior distributions on the posterior distribution in Bayesian inference. Exploring the posteriors under several alternative priors can be computationally intensive, particularly for complex latent variable models. To address this issue, we propose a novel method for quantifying the prior sensitivity that does not require model re-fit. Specifically, we present a method to compute the Hellinger and Kullback-Leibler distances between two posterior distributions with base and alternative priors, using Monte Carlo integration based only on the base posterior distribution, through novel integral expressions of the two distances. We also extend the above approach for assessing the influence of hyperpriors in general latent variable models. We demonstrate the proposed method through examples of a simple normal distribution model, hierarchical binomial-beta model, and Gaussian process regression model.

\bigskip\noindent
{\bf Key words}: Kullback Leibler divergence; Hellinger divergence; marginal likelihood; Monte Carlo integration

\newpage
\section{Introduction}

In Bayesian inference, the choice of prior distributions representing beliefs or knowledge before observing data may influence posterior distributions.
As robustness to prior distributions ensures that conclusions are not affected by subjective or arbitrary prior choice, prior sensitivity analysis is a critical step in the Bayesian workflow \citep[e.g.][]{depaoli2020importance,gelman2020bayesian}.
Regarding the theoretical argument of the prior sensitivity analysis, the classical discussions on prior sensitivity by \citep{berger1986robust} and \cite{berger1990robust} have triggered various studies from theoretical perspectives \citep[e.g.][]{kurtek2015bayesian}.
However, these papers are not concerned with the efficient computation of measures for prior sensitivity. 
A typical way of sensitivity analysis is to re-fit the model with different prior distributions and investigate the difference in characteristics such as posterior means and variances.
That being said, such an approach requires substantial computation time \citep[e.g.][]{perez2006mcmc}, especially when the number of parameters is large, or the model includes a large number of latent variables. 
While there are computationally efficient approaches in terms of computation time, most of them are designed for specific models \citep[e.g.][]{van2018prior,jackson2019value,roos2021sensitivity,sona2023quantification,giordano2023evaluating}, and cannot be applied to a general model.

To address these challenges, we propose a novel and general method for prior sensitivity analysis.
A key to our approach is that the proposed method does not require model re-fitting, which significantly reduces the computational costs in prior sensitivity analysis.
As measures of prior sensitivity, we propose an algorithm to compute the Hellinger and Kullback-Leibler distances between two posterior distributions based on the base and alternative priors, using only Monte Carlo integration from the base posterior distribution. 
Furthermore, we derive a formula to calculate posterior expectations under the alternative prior, based only on the base posterior. This enables us to investigate the effect of prior distribution on various characteristics of the alternative posterior. 
Owing to a significant reduction of the computational cost to evaluate alternative posteriors, one can carry out a more comprehensive evaluation of alternative priors. 
The proposed method is not model specific such that it will be adopted in arbitrary Bayesian models and be used with a general computation tool for Bayesian inference \citep[e.g.][]{vstrumbelj2024past}.
Moreover, we also consider the same argument in hierarchical or latent variable models, for which prior sensitivity is particularly essential to determine which model components are difficult to learn from the data.
For general latent variable models, we provide Monte Carlo approximation for measuring sensitivity regarding joint posterior of model parameters and latent variables and marginal posteriors of latent variables, which facilitates a wide range of prior sensitivity analysis for complex latent variable models.

There is a substantial body of research on efficient methods for prior sensitivity analysis. 
\cite{kass1989approximate} and \cite{kass1992approximate} proposed the use of Laplace approximation for computing sensitivity measures, but precise evaluation cannot be achieved due to its approximation nature.
\cite{kallioinen2024detecting} introduced a method for assessing the sensitivity of priors and likelihoods using power-scaling. Still, the class of alternative priors it can handle is limited, and importance sampling may render it impractical in high-dimensional parameter settings. 
For prior sensitivity analysis in latent variable models, \cite{roos2015sensitivity} proposed a method without model re-fitting. 
Nonetheless, it relies on numerical integration using density estimation of the posterior distribution, which could be practically difficult when the parameter dimension is not small.

The organization of this paper is as follows. 
Section~\ref{sec:main} introduces our proposed Monte Carlo methods for computing H-distance, KL-distance, and posterior expectations under posteriors with alternative priors for a general statistical model. 
We also provide extensions to general latent variables models. 
In Section~\ref{sec:example}, we demonstrate the proposed sensitivity methods through three numerical examples.
Concluding remarks are given in Section~\ref{sec:conc}, and all the proofs are deferred in the Appendix. 
The R codes to replicate the results in this paper are available at GitHub repository (\url{https://github.com/sshonosuke/prior_sensitivity}).

\section{Prior Sensitivity Analysis without Model Re-fit}\label{sec:main}

\subsection{Sensitivity measures and their computation}
Let $f(x|\theta)$ be a likelihood function with parameter $\theta$ and data $x$, and $\pi(\theta)$ be a (base) prior distribution for $\theta$. 
The posterior distribution of $\theta$ is given by 
\begin{equation}\label{eq:pos}
\pi(\theta|x)=\frac{\pi(\theta)f(x|\theta)}{\int\pi(\theta)f(x|\theta)d\theta}.
\end{equation}
To check the sensitivity of the posterior (\ref{eq:pos}) to the choice of the prior distribution, we are interested in how the posterior $\pi_{\ast}(\theta|x)$ with a different prior $\pi_{\ast}(\theta)$ differs from the base posterior (\ref{eq:pos}).
To compare two posterior distributions, we employ two measures, Kullbuck-Leibler (KL) and Hellinger (H) distance, which are respectively defined as 
\begin{equation}\label{eq:FR}
\begin{split}
{\rm KL}(\pi(\theta|x)\|\pi_{\ast}(\theta|x))
&\equiv 
\int \pi(\theta|x) \log\left\{\frac{\pi(\theta|x)}{ \pi_{\ast}(\theta|x)}\right\}d\theta,  \\
{\rm H}( \pi(\theta|x)\|\pi_{\ast}(\theta|x) )^2 
&\equiv \frac12\int \left(\sqrt{\pi(\theta|x)}-\sqrt{\pi_{\ast}(\theta|x)}\right)^2 d\theta.
\end{split}
\end{equation}
The second-order approximation of the Hellinger distance is known to be equivalent to the Fisher-Rao metric when the two distributions are close to each other \citep[e.g.][]{amari2000methods}.
The Fisher-Rao metric is a natural distance measure on the manifold consisting of statistical models and has emerged as a standard measure for evaluating sensitivity and robustness of posterior \citep[e.g.][]{kurtek2015bayesian}.

By using the two sensitivity measures (H-sensitivity and KL-sensitivity), we can quantify the prior effect as a value on $[0, 1]$ for the H-sensitivity and $[0, \infty)$ for the KL-sensitivity.
However, the crucial problem with using the metric (\ref{eq:FR}) is that their computation under general models is not straightforward, where generating posterior samples from $\pi_{\ast}(\theta|x)$ for a variety of alternative prior distributions would be computationally demanding. 
In the following theorem, we propose an alternative representation of the two metrics in (\ref{eq:FR}) without computing the alternative posterior $\pi_{\ast}(\theta|x)$, namely, by using only the base posterior distribution $\pi(\theta|x)$.

\begin{thm}\label{thm1}
The H-sensitivity and KL-sensitivity between the base and alternative posterior distributions in (\ref{eq:FR}) are expressed as 
\begin{equation}\label{eq:score}
\begin{split}
{\rm H}( \pi(\theta|x)\|\pi_{\ast}(\theta|x) )^2 
&=1- E_{\pi(\theta|x)}\left[\sqrt{\frac{\pi_{\ast}(\theta)}{\pi(\theta)}}\right] \bigg/ \sqrt{E_{\pi(\theta|x)}\left[\frac{\pi_{\ast}(\theta)}{\pi(\theta)}\right]},\\
{\rm KL}(\pi(\theta|x)\|\pi_{\ast}(\theta|x))
&=  \log \left\{E_{\pi(\theta|x)}\left[\frac{\pi_{\ast}(\theta)}{\pi(\theta)}\right]\right\}- E_{\pi(\theta|x)}\left[\log \frac{\pi_{\ast}(\theta)}{\pi(\theta)}\right].
\end{split}
\end{equation}
where $E_{\pi(\theta|x)}$ denotes the expectation with respect to the base posterior distribution $\pi(\theta|x)$.
\end{thm}

The formulae (\ref{eq:score}) can be computed by the posterior expectation of the ratios of two prior distributions with respect to the base posterior $\pi(\theta|x)$.
Hence, these values can be easily computed by posterior samples from $\pi(\theta|x)$ obtained by, for example, the Markov Chain Monte Carlo algorithm.  
This drastically reduces the computation time as it does not require model re-fit with alternative priors.  
Further, the above expression is free from density estimation of the base posterior or importance (re-)sampling, as required in the existing approaches \citep[e.g.][]{roos2015sensitivity,kallioinen2024detecting}, which can be computationally prohibitive when the dimension of $\theta$ is large.

While the integral representation (\ref{eq:score}) entails integral in $D$ (dimension of $\theta$) dimension, the essential dimension could be small, depending on the settings of alternative priors.
For example, suppose that a base prior for $\theta=(\theta_1,\ldots,\theta_D)$ is independent, $\pi(\theta)=\prod_{d=1}^D\pi(\theta_d)$, and consider an alternative prior by modifying only the prior for $\theta_1$ while keeping the others unchanged, namely, $\pi_{\ast}(\theta)=\pi_{\ast}(\theta_1)\prod_{d=2}^D\pi(\theta_d)$.
Then, the expression for the H-sensitivity is reduced to 
$$
{\rm H}( \pi(\theta|x)\|\pi_{\ast}(\theta|x) )^2 
=1- E_{\pi(\theta_1|x)}\left[\sqrt{\frac{\pi_{\ast}(\theta_1)}{\pi(\theta_1)}}\right] \bigg/ \sqrt{E_{\pi(\theta_1|x)}\left[\frac{\pi_{\ast}(\theta_1)}{\pi(\theta_1)}\right]},
$$
which only requires one-dimensional integration, where $\pi(\theta_1|x)$ is the marginal posterior of $\theta_1$.  
On the other hand, existing approaches using kernel density estimation of posteriors needs $D$-dimensional integration regardless of the form of alternative priors.

It is important to carefully consider the difference in the spread between the alternative prior and the base prior. In general, if the alternative prior has a wider spread than the base prior, Monte Carlo integration may become unstable.
As a general recommendation, it is preferable to set a distribution with a larger spread as the base prior and consider a distribution with a smaller spread as the alternative prior. This approach helps ensure more stable Monte Carlo estimation.

\subsection{Prior sensitivity analysis in latent variables models}\label{sec:main-LVM}
We next consider prior sensitivity on the posterior distribution of latent variables.
We consider the following general latent variables model can be expressed as 
\begin{equation}\label{eq:LVM}
x|\eta,\psi\sim f(x|\eta,\psi), \ \ \ \eta|\theta\sim p(\eta|\theta),
\end{equation}
where $\eta$ is a latent variable, $\psi$ is a parameter in the conditional distribution of the observed data $x$ given $\eta$, and $\theta$ is a parameter in the distribution of the latent variable $\eta$.
The model (\ref{eq:LVM}) contains various hierarchical or latent variables models (e.g., generalized linear mixed models and state space models) adopted in practice. 

Let $\pi(\psi,\theta)$ be a base prior for $(\psi,\theta)$.
Then, the joint posterior of the latent variable $\eta$ and model parameters $(\psi,\theta)$ is given by 
$$
\pi(\eta, \psi,\theta|x) =\frac{f(x|\eta,\psi)p(\eta|\theta)\pi(\psi,\theta)}{m(x)},
$$
where $m(x)$ is the marginal likelihood given by 
$$
m(x)=\int f(x|\eta,\psi)p(\eta|\theta)\pi(\psi,\theta)d\eta d\psi d\theta.
$$
We define the alternative joint posterior $\pi_{\ast}(\eta, \psi,\theta|x)$ as one obtained by replacing $\pi(\psi,\theta)$ with $\pi_{\ast}(\psi,\theta)$ in the above expression. 
We also define $\pi(\psi,\theta|x)=\int \pi(\eta,\psi,\theta|x)d\eta$ as the marginal posterior of $(\psi,\theta)$.
We first show that the H-sensitivity and KL-distance between $\pi(\eta, \psi,\theta|x)$ and $\pi_{\ast}(\eta, \psi,\theta|x)$ can be expressed as the posterior expectation with respect to the base posterior $\pi(\eta, \psi,\theta|x)$, as shown in the following theorem.

\begin{thm}
The H-sensitivity and KL-sensitivity for the two joint posterior distributions of latent variables and parameters under different priors are expressed as 
\begin{equation}\label{eq:D2}
\begin{split}
{\rm H}(\pi(\eta,\psi,\theta|x) \| &\pi_{\ast}(\eta,\psi,\theta|x))^2\\
&=1-E_{\pi(\psi,\theta|x)}\left[\sqrt{\frac{\pi_{\ast}(\psi,\theta)}{\pi(\psi,\theta)}}\right] \bigg/ \sqrt{E_{\pi(\psi,\theta|x)}\left[\frac{\pi_{\ast}(\psi,\theta)}{\pi(\psi,\theta)}\right]},\\
{\rm KL}(\pi(\eta,\psi,\theta|x) \| &\pi_{\ast}(\eta,\psi,\theta|x))\\
&=\log\left\{E_{\pi(\psi,\theta|x)}\left[\frac{\pi_{\ast}(\psi,\theta)}{\pi(\psi,\theta)}\right]\right\} 
- E_{\pi(\psi,\theta|x)}\left[\log \left(\frac{\pi_{\ast}(\psi,\theta)}{\pi(\psi,\theta)}\right)\right].
\end{split}
\end{equation}
\end{thm}

\vspace{0.5cm}
The above formulae can be readily computed through posterior samples of $(\psi, \theta)$ from the base posterior, which again can be done by MCMC. 
While the formula (\ref{eq:D2}) is for the joint posteriors, one may be interested in sensitivity only on the marginal posteriors of the latent variable $\eta$. 
The marginal posterior of $\eta$ is given by 
$$
p(\eta| x)=\frac1{m(x)}\int f(x|\eta,\psi)p(\eta|\theta)\pi(\psi,\theta)d\psi d\theta.
$$
We then define $\pi_{\ast}(\eta|x)$ as the marginal posterior of $\eta$ with an alternative prior $\pi_{\ast}(\psi,\theta)$ instead of $\pi(\psi,\theta)$. 
We then have the following results for the distance between the base and alternative posteriors.

\begin{thm}
The H-sensitivity and KL-sensitivity for the two marginal posterior distributions of latent variables $\eta$ under different priors for $(\psi,\theta)$ are expressed as 
\begin{align*}
{\rm H}(\pi(\eta|x) &\| \pi_{\ast}(\eta|x))^2\\
&=1- E_{\pi(\eta|x)}\left[\sqrt{E_{\pi(\psi,\theta|\eta,x)}\left[\frac{\pi_{\ast}(\psi,\theta)}{\pi(\psi,\theta)}\right]}\right] \bigg/ \sqrt{E_{\pi(\psi,\theta|x)}\left[\frac{\pi_{\ast}(\psi,\theta)}{\pi(\psi,\theta)}\right]}, \\
{\rm KL}(\pi(\eta|x) &\| \pi_{\ast}(\eta|x))\\
&=\log\left\{E_{\pi(\psi,\theta|x)}\left[\frac{\pi_{\ast}(\psi,\theta)}{\pi(\psi,\theta)}\right]\right\} 
- E_{\pi(\eta|x)}\left[\log \left(E_{\pi(\psi,\theta|\eta,x)}\left[\frac{\pi_{\ast}(\psi,\theta)}{\pi(\psi,\theta)}\right]\right)\right].
\end{align*}
\end{thm}

\medskip
Again, the above formula does not depend on the alternative posterior $\pi_{\ast}(\eta,\psi,\theta|x)$, so it does not require model re-fit. 
The above formula includes the expectation of the form $E_{\pi(\eta|x)}\big[\gamma(E_{\pi(\psi,\theta|\eta,x)}[\alpha(\psi,\theta)])\big]$ for some functions $\gamma(\cdot)$ and $\alpha(\cdot, \cdot)$. 
Note that the inner expectation is about the conditional posterior of $(\psi,\theta)$ given $(\eta,x)$, which cannot be directly obtained from the Monte Carlo sample from the joint posterior $\pi(\eta,\psi,\theta|x)$.
Here, we propose computing the expectation as 
$$
E_{\pi(\eta|x)}\big[\gamma(E_{\pi(\psi,\theta|\eta,x)}[\alpha(\psi,\theta)])\big]
\approx \frac1S\sum_{s=1}^{S}\gamma\bigg(\frac{1}{|I({\eta^{(s)}})|}\sum_{r\in I({\eta^{(s)}})}\alpha(\psi^{(r)}, \theta^{(r)})\bigg),
$$
where $\{\eta^{(s)},\psi^{(s)}, \theta^{(s)}\}_{s=1,\ldots,S}$ are the Monte Carlo sample from the base posterior, and $I(\eta)=\{r\in \{1,\ldots,S\} \ | \ \|\eta^{(r)}-\eta\|<\varepsilon\}$ for some small $\varepsilon>0$.

\section{Illustrations}\label{sec:example}

\subsection{Normal distribution}
\begin{figure}[htbp!]
\centering
\includegraphics[width=\linewidth]{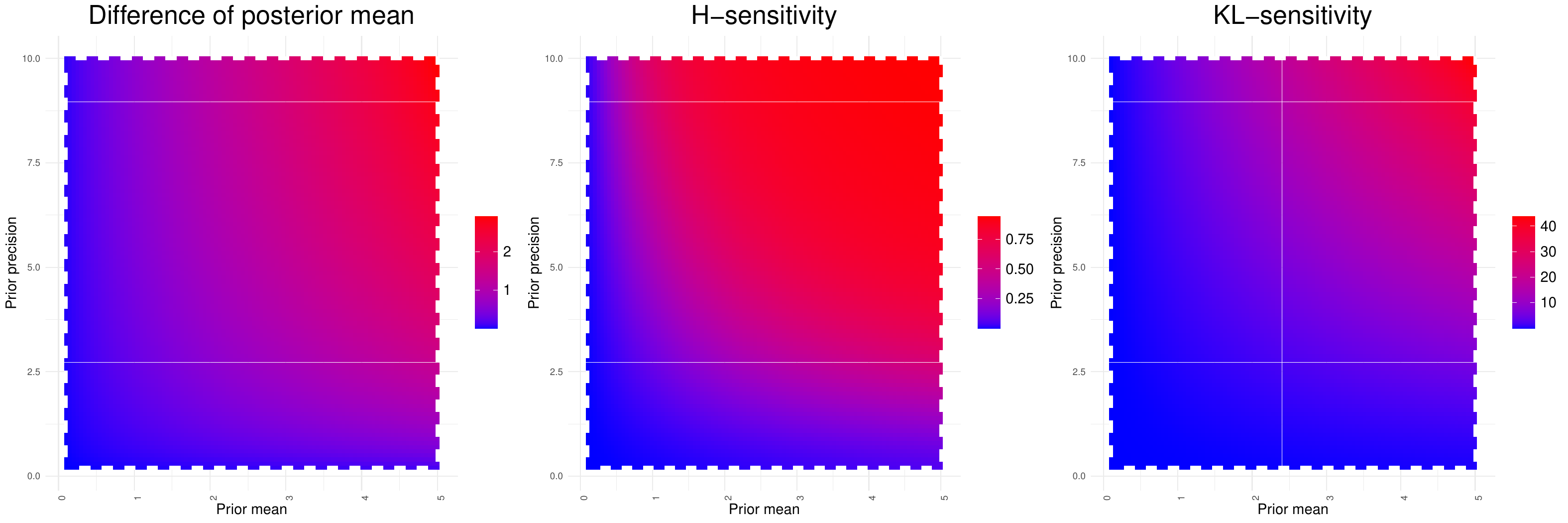}
\caption{Difference of posterior mean (left) and the H-sensitivity (right) under alternative priors. } 
\label{fig:normal}
\end{figure}

We first consider a simple example using normal distributions with unknown mean parameters to show that the H-sensitivity could be used to investigate the prior sensitivity.
For observations, $x_1,\ldots,x_n$, we apply the normal distribution, $N(\mu, 1)$, with prior $\mu\sim N(\mu_0, \tau_0^{-1})$ with fixed hyperparameters $\mu_0$ and $\tau_0$.
Under the prior, the posterior distribution of $\mu$ is given by $\mu|x\sim N((n+\tau_0)^{-1}(n\bar{x} + \tau_0\mu_0), (n+\tau_0)^{-1})$, where $\bar{x}=n^{-1}\sum_{i=1}^n x_i$.
Suppose that we observe $(x_1,\ldots,x_7)=(-2, -1, -0.5, 0, 0.5, 1, 2)$, consider the base prior with $\mu_0=0$ and $\tau_0=10^{-4}$.
Then, we apply alternative choices of hyperparameters and evaluate the H-sensitivity and KL-sensitivity between the two posteriors based on the base and alternative priors. 
In the left panel of Figure~\ref{fig:normal}, we show the posterior mean of $\mu$ under various hyperparameters. 
It can be observed that the posterior mean of the alternative posterior is very different from that of the base posterior when the prior mean is away from 0, and the prior precision is large. 
In the center and right panel of Figure~\ref{fig:normal}, we present the two sensitivity measures between the base and alternative posteriors, using Theorem~1 with 1000 Monte Carlo samples of the base posterior. 
It shows that the two sensitivity measures increase according to the difference of posterior means, which indicates that the both measures can successfully detect the disparity from the base posterior.

\subsection{Hierarchical binomial-beta models}

We next illustrate an example of latent variable models with different parameterizations. 
Let $y_i \ (i=1,\ldots,m)$ be the number of successes among $n_i$ trials, for which we consider the following hierarchical binomial-beta model: 
$$
y_i|\theta_i\sim {\rm Bin}(n_i, \theta_i), \ \ \ \ \theta_i\sim {\rm Beta}(\alpha, \beta), \ \ \ \ i=1,\ldots,m,
$$
where ${\rm Bin}(n_i, \theta_i)$ is a binomial distribution with probability $\theta_i$ and ${\rm Beta}(\alpha, \beta)$ is a beta distribution with parameter $\alpha$ and $\beta$. 
In this model, $\theta_1,\ldots,\theta_m$ are latent variables and $(\alpha, \beta)$ are model parameters for the prior distribution of $\theta_i$.
For fitting the model, we consider the following parameterization of the beta distribution:
$$
\delta=-\log\left(\frac{\alpha}{\alpha+\beta}\right),\ \ \ \gamma=\frac{1}{\sqrt{\alpha+\beta}}.
$$
Note that the above transformation is one-to-one, and both $\delta$ and $\gamma$ are positive valued parameters as $\alpha$ and $\beta$.
Since $E[\theta_i]=\exp(-\delta)$ and ${\rm Var}(\theta_i)=E[\theta_i](1-E[\theta_i])\gamma^2/(1+\gamma^2)$, $\delta$ and $\gamma$ control prior mean and scale of $\theta_i$, respectively. 
The parameterization using $(\delta, \gamma)$ and $(\alpha, \beta)$ will be denoted by ``Parameterization 1" and ``Parameterization 2", respectively.

\begin{figure}[t]
\centering
\includegraphics[width=0.8\linewidth]{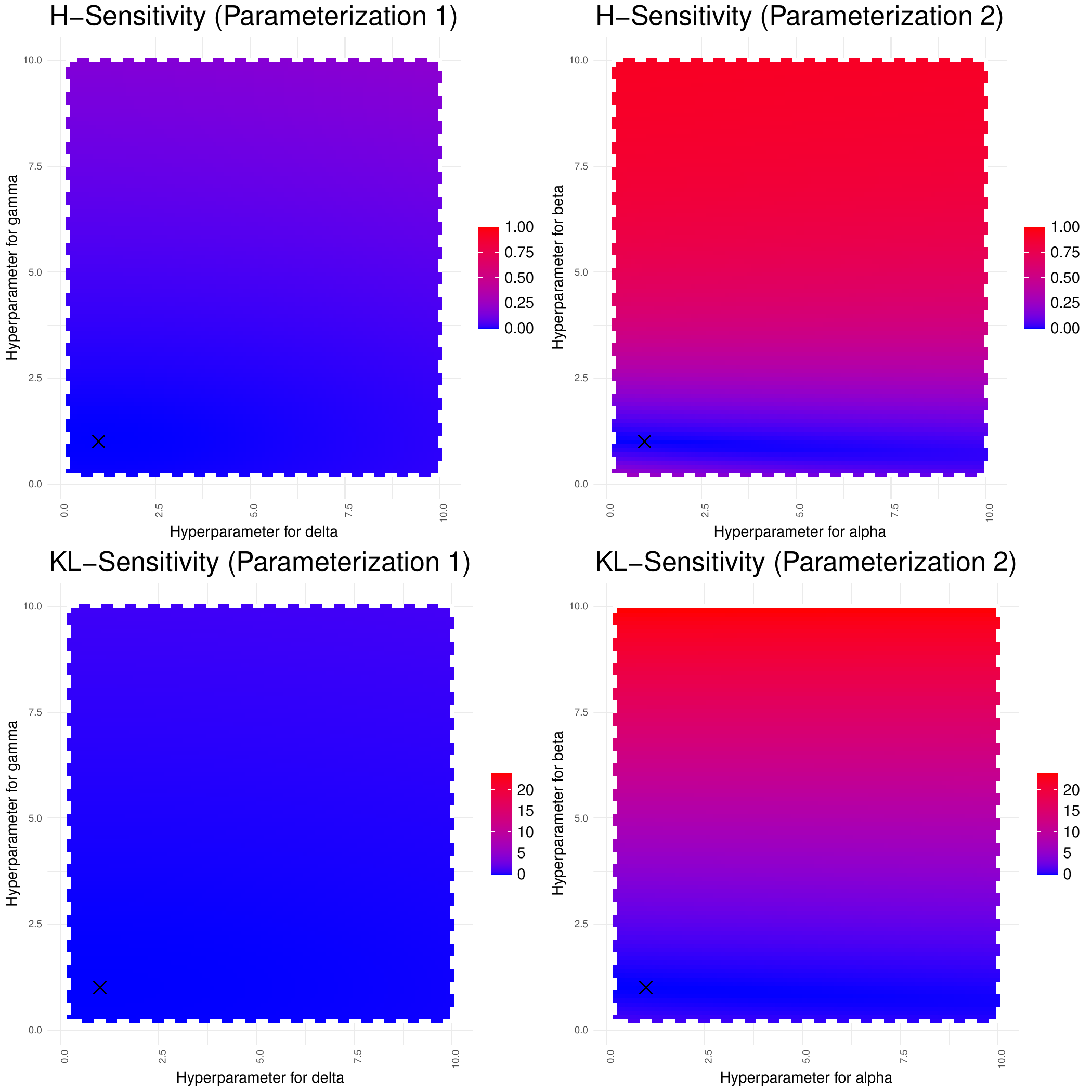}
\caption{The H-sensitivity between the baseline and alternative priors under parameterization 1 (left) and parameterization 2 (right) of the binomial-beta model. } 
\label{fig:BB}
\end{figure}

For illustration, we employ rat tumor data \citep[e.g., Chapter xx, ][]{gelman2013bayesian}, which includes the number of rats having tumors, $y_i$ among $n_i$ rats in $m=71$ groups. 
For each parameterization, we first fit the binomial-beta model with base priors, $\delta,\gamma\sim {\rm Ga}(1,1)$ (Parameterization 1) and $\alpha,\beta\sim {\rm Ga}(1,1)$ (Parameterization 2), implemented by using \verb+Stan+ \citep{carpenter2017stan}. 
For evaluating the prior sensitivity, we consider a class of alternative priors, ${\rm Ga}(\nu, \nu)$, with $\nu\in (0, 10]$, for $\delta, \gamma, \alpha$ and $\beta$. 
Using Theorem~3 and 4000 posterior samples after discarding 4000 samples as burn-in, we computed the H-sensitivity and KL-sensitivity between the base and alternative posteriors. 
Figure~\ref{fig:BB} shows the results for two parameterizations.
First, both sensitivity measures detect similar patterns of prior sensitivity.  
It can also be seen that in Parameterization 2, changing the prior leads to sensitive variations in the posterior distribution, whereas in Parameterization 1, the posterior is not sensitive to the prior. This is because, in Parameterization 2, extracting information for $(\alpha,\beta)$ from the data is difficult, making the posterior heavily influenced by the prior distribution.

\subsection{Gaussian process regression}

We here employ Gaussian process regression to illustrate the sensitivity measures for joint as well as marginal posteriors. 
Let $y_i \ (i=1,\ldots,n)$ be a response variable and $x_i$ be a scalar covariate.
Then, the Gaussian process regression model is described as 
\begin{align*}
&y_i=f(x_i)+\varepsilon_i, \ \ \ \varepsilon_i\sim N(0, \sigma^2), \ \ \ \ i=1,\ldots,n, \\
&\ \  (f(x_1),\ldots,f(x_n)) \sim N(0, C_n(\tau, \psi)),
\end{align*}
where $\sigma^2$ is an error variance and $C_n(\tau, \psi)$ is an $n\times n$ covariate matrix whose $(i,j)$-element being $\tau^2\rho(|x_i-x_j|; \psi)$ for some valid correlation function $\rho(\cdot; \psi)$ characterized by the parameter $\psi$. 
In this study, we will use the exponential correlation function, $\rho(d; \psi)=\exp(-d/\psi)$, where $\psi$ is known as the range parameter.

\begin{figure}[t]
\centering
\includegraphics[width=0.8\linewidth]{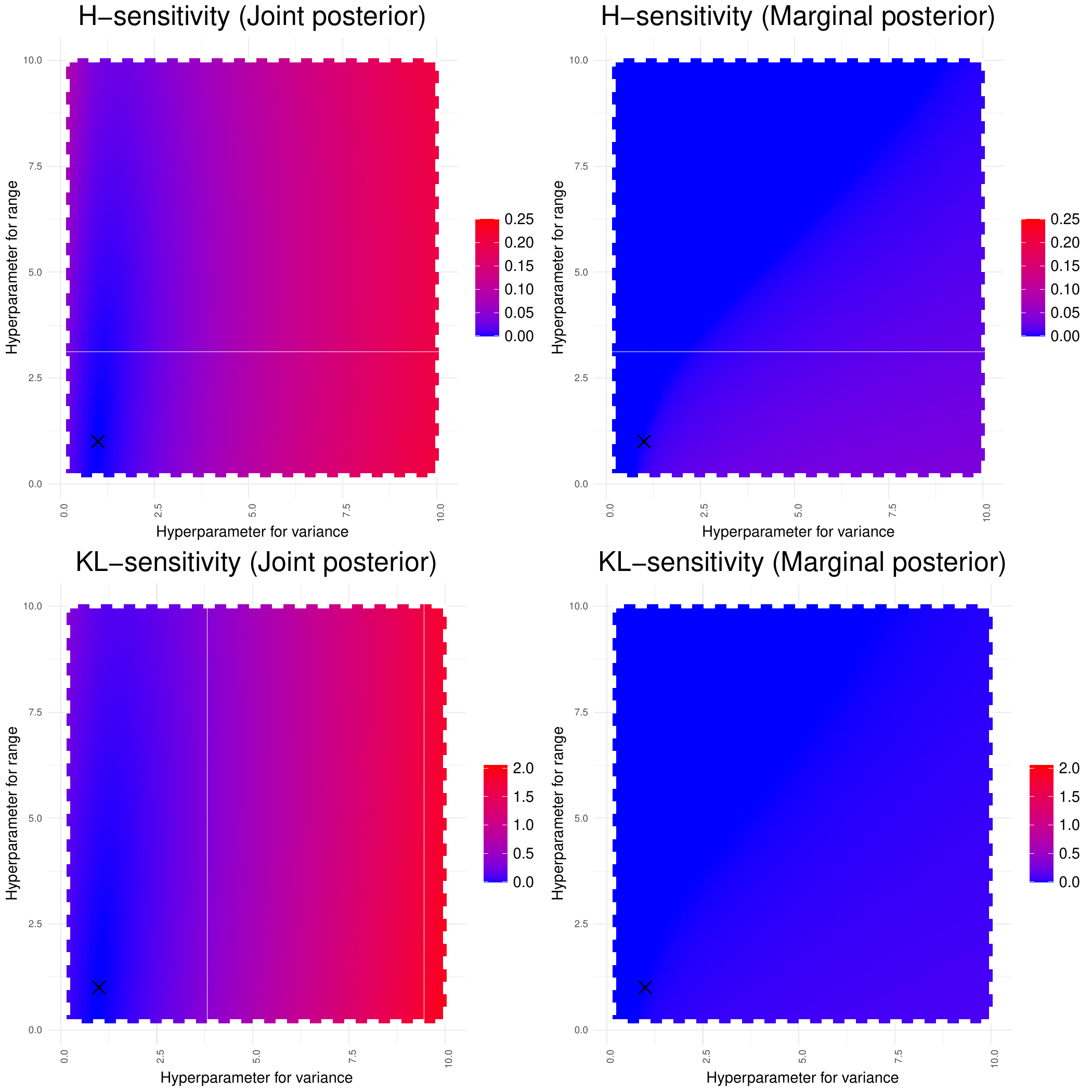}
\caption{H-sensitivity for the joint posterior of latent variables and parameters (left) and marginal posterior of latent variables (right). The hyperparameters of the base posterior are marked with a cross. } 
\label{fig:GP}
\end{figure}

For illustration, we generated synthetic data $\{x_i,y_i\}_{i=1,\ldots,n}$ with $n=50$ from 
$$
y_i=\sin(\pi x_i) + x_i  + \varepsilon_i, \ \ \ x_i\sim U(0,3), \ \ \ \varepsilon_i\sim N(0, (0.5)^2).
$$
We implemented the Gaussian process model by \verb+Stan+ with prior for $\sigma^2, \tau^2, \psi$ being ${\rm Ga}(1,1)$, and generated 1000 posterior samples after discarding the first 1000 samples as burn-in.
To check the prior sensitivity on $\tau^2$ and $\psi$ (Gaussian process parameters), we consider a class of alternative priors, $\tau^2\sim {\rm Ga}(\delta_\tau, \delta_\tau)$ and $\psi \sim {\rm Ga}(\delta_\psi, \delta_\psi)$ for $\delta_\tau, \delta_\psi\in (0, 10]$. 
Using Theorem~3 and 4, we calculated the H-sensitivity and KL-sensitivity of the joint posterior as well as the marginal posterior of latent variables, $(f(x_1),\ldots,f(x_n))$, where the results are shown in Figure~\ref{fig:GP}.
It is observed that the two sensitivity measures detect similar patterns of prior sensitivity, as in the previous example. 
We can also see that the joint posterior exhibits greater sensitivity compared to the marginal posterior of the latent variables. 
Moreover, the prior on the scale parameter $\tau$ has significant influence on the posterior, suggesting that a prior for $\tau$ might be carefully chosen. 
On the other hand, the posterior distribution of the latent variables is not directly affected by the change of prior distributions.

\section{Concluding Remarks}\label{sec:conc}

We developed a general and convenient method for evaluating prior sensitivity without model re-fit, where the sensitive measures can be computed by only using the base posterior. 
We also present methods for sensitivity analysis of latent variables models, which facilitates comprehensive assessment of the sensitivity under a wide range of priors.    
While this paper is focused on sensitivity with respect to prior distributions, likelihood sensitivity might also be of interest. 
It would be valuable to consider general formulae of sensitivity when both prior and likelihood are changed as considered in \cite{kallioinen2024detecting}.
Moreover, the proposed method would also be useful in evaluating $\varepsilon$-local sensitivity \citep[e.g.][]{mcculloch1989local,zhu2011bayesian}, which will be a valuable future study.

\section*{Acknowledgement}
This work is supported by the Japan Society for the Promotion of Science (grant numbers: 20H00080, 21H00699, 24K00244, 24K00175, and 24K21420).

\newpage
\appendix
\setcounter{equation}{0}
\renewcommand{\theequation}{A\arabic{equation}}

\vspace{0.5cm}
\begin{center}
{\bf {\large Appendix}}
\end{center}

\section*{A1. Proof of Theorem~1}
First, we note that 
$$
{\rm H}( \pi(\theta|x)\|\pi_{\ast}(\theta|x) )^2 
= 1-\int \sqrt{\pi(\theta|x)} \sqrt{\pi_{\ast}(\theta|x)} d\theta.
$$
Then, it follows that 
\begin{align*}
\int \sqrt{\pi(\theta|x)} \sqrt{\pi_{\ast}(\theta|x)} d\theta
&=\int \pi(\theta|x) \left(\frac{\pi_{\ast}(\theta|x)}{\pi(\theta|x)}\right)^{1/2} d\theta\\
&=\int \pi(\theta|x) \left(\frac{f(x|\theta)}{f(x|\theta)}\frac{\pi_{\ast}(\theta)}{\pi(\theta)}\frac{m(x)}{m_{\ast}(x)}\right)^{1/2} d\theta\\
&=\left(\frac{m_{\ast}(x)}{m(x)}\right)^{-1/2}E_{\pi(\theta|x)}\left[\left(\frac{\pi_{\ast}(\theta)}{\pi(\theta)}\right)^{1/2}\right], 
\end{align*}
where $m(x)=\int\pi(\theta)f(x|\theta)d\theta$ and $m_{\ast}(x)=\int\pi_{\ast}(\theta)f(x|\theta)d\theta$ are the marginal likelihood under the base prior $\pi(\theta)$ and alternative prior $\pi_{\ast}(\theta)$, respectively.
Furthermore, the likelihood ratio, $m(x)/m_{\ast}(x)$, can be written as 
\begin{equation}\label{eq:ML-ratio}
\frac{m_{\ast}(x)}{m(x)}
=\int \frac{f(x|\theta)\pi_{\ast}(\theta)}{m(x)}d\theta
=\int \pi(\theta|x) \frac{f(x|\theta)\pi_{\ast}(\theta)}{f(x|\theta)\pi(\theta)}d\theta 
=E_{\pi(\theta|x)}\left[\frac{\pi_{\ast}(\theta)}{\pi(\theta)}\right],
\end{equation}
where we used the expression of the marginal likelihood, $m(x)=f(x|\theta)\pi(\theta)/\pi(\theta|x)$.
This completes the derivation for the H-sensitivity in (\ref{eq:score}).

Regarding the KL-sensitivity, it holds that 
\begin{align*}
\int \pi(\theta|x) \log\left\{\frac{\pi(\theta|x)}{\pi_{\ast}(\theta|x)}\right\}d\theta
&=
\int \pi(\theta|x) \log\left\{\frac{f(x|\theta)}{f(x|\theta)}\frac{\pi(\theta)}{\pi_{\ast}(\theta)}\frac{m_{\ast}(x)}{m(x)}\right\}d\theta\\
&=
\log\left\{\frac{m_{\ast}(x)}{m(x)}\right\} - 
E_{\pi(\theta|x)}\left[\log\left(\frac{\pi_{\ast}(\theta)}{\pi(\theta)}\right)\right]. 
\end{align*}
Hence, from (\ref{eq:ML-ratio}), we obtain the formula given in (\ref{eq:score}).

\section*{A2. Proof of Theorem~2}
It follows that 
\begin{align*}
\int & \sqrt{\pi(\eta,\psi,\theta|x)} \sqrt{\pi_{\ast}(\eta,\psi,\theta|x)} d\eta d\psi d\theta\\
&=\int p(\eta,\psi,\theta|x)\left(
\frac{f(x|\eta,\psi)p(\eta|\theta)\pi(\psi,\theta)/ m(x)}
{f(x|\eta,\psi)p(\eta|\theta)\pi_{\ast}(\psi,\theta)/ m_{\ast}(x) }
\right)^{1/2} d\eta d\psi d\theta \\
&=\left(\frac{m_{\ast}(x)}{m(x)}\right)^{-1/2}
\int \pi(\eta,\psi,\theta|x)\left(
\frac{\pi(\psi,\theta) }
{\pi_{\ast}(\psi,\theta)}
\right)^{1/2} d\eta d\psi d\theta,
\end{align*}
noting that the last integral can be expressed as an expectation with respect to the marginal posterior, $\pi(\psi,\theta|x)$.
Regarding the ratio, $m_{\ast}(x)/m(x)$, it holds that 
\begin{equation}\label{eq:ML-ratio2}
\begin{split}
\frac{m_{\ast}(x)}{m(x)}
&=\frac{1}{m(x)}\int f(x|\eta,\psi)p(\eta|\theta)\pi_{\ast}(\psi,\theta) d\eta d\psi d\theta\\
&=
\int \pi(\eta,\psi,\theta|x) \frac{ f(x|\eta,\psi)p(\eta|\theta)\pi_{\ast}(\psi, \theta)}
{ f(x|\eta,\psi)p(\eta|\theta)\pi(\psi, \theta)}d\eta d\theta d\psi\\
&=
\int \pi(\psi,\theta|x) \frac{ \pi_{\ast}(\psi, \theta)}
{ \pi(\psi, \theta)} d\theta d\psi,
\end{split}
\end{equation}
which establishes the result for the H-sensitivity between $\pi(\eta,\psi,\theta|x)$ and $\pi_{\ast}(\eta,\psi,\theta|x)$.

For the KL-sensitivity, it follows that 
\begin{align*}
\int &\pi(\eta,\psi,\theta|x)  \log \left\{\frac{\pi(\eta,\psi,\theta|x) }{\pi_{\ast}(\eta,\psi,\theta|x) }\right\}d\eta d\psi d\theta\\
&=
\int \pi(\eta,\psi,\theta|x)\log\left\{\frac{f(x|\eta,\psi)}{f(x|\eta,\psi)}\frac{p(\eta|\theta)}{p(\eta|\theta)}\frac{\pi(\psi,\eta)}{\pi_{\ast}(\psi,\eta)}\frac{m_{\ast}(x)}{m(x)}\right\}d\theta\\
&=
\log\left\{\frac{m_{\ast}(x)}{m(x)}\right\} - 
E_{\pi(\psi,\theta|x)}\left[\log\left(\frac{\pi_{\ast}(\psi,\theta)}{\pi(\psi,\theta)}\right)\right].
\end{align*}
Using (\ref{eq:ML-ratio2}), we obtain the result for the KL-sensitivity.

\section*{A3. Proof of Theorem~3}

The H-sensitivity for the marginal posterior of $\eta$ can be expressed as 
\begin{align*}
\int \sqrt{\pi(\eta|x)}&\sqrt{\pi_{\theta\ast}(\eta|x)}d\eta\\
&=\int \pi(\eta|x)\left(
\frac{\int f(x|\eta,\psi)\pi(\eta|\theta)\pi_{\ast}(\psi,\theta)d\psi d\theta / m_{\ast}(x)}
{\int f(x| \eta,\psi)\pi(\eta|\theta)\pi(\psi,\theta)d\psi d\theta / m(x) }
\right)^{1/2} d\eta \\
&=\left(\frac{m_{\ast}(x)}{m(x)}\right)^{-1/2}
\int \pi(\eta|x)\left(
\frac{\int f(x|\eta,\psi)\pi(\eta|\theta)\pi_{\ast}(\psi,\theta)d\psi d\theta}
{\int f(x| \eta,\psi)\pi(\eta|\theta)\pi(\psi,\theta)d\psi d\theta }
\right)^{1/2}d\eta,
\end{align*}
where the ratio of the two marginal likelihood $m_{\ast}(x)/m(x)$ can be rewritten as (\ref{eq:ML-ratio2}).
Furthermore, since the conditional posterior of $(\psi,\theta)$ given $(\eta, x)$ is expressed as
$$
\pi(\psi,\theta|\eta, x)=\frac{f(x|\eta,\psi)\pi(\eta|\theta)\pi(\psi,\theta)}{\int f(x|\eta,\psi)\pi(\eta|\theta)\pi(\psi,\theta)d\psi d\theta},
$$
it holds that 
\begin{equation}\label{eq:ML-ratio4}
\begin{split}
\frac{\int f(x|\eta,\psi)\pi(\eta|\theta)\pi_{\ast}(\psi,\theta)d\psi d\theta}
{\int f(x| \eta,\psi)\pi(\eta|\theta)\pi(\psi,\theta)d\psi d\theta }
&=
\int \pi(\psi,\theta|\eta, x) \frac{f(x|\eta,\psi)\pi(\eta|\theta)\pi_{\ast}(\psi,\theta)}{f(x|\eta,\psi)\pi(\eta|\theta)\pi(\psi,\theta)}d\psi d\theta\\
&=\int \pi(\psi,\theta|\eta, x) \frac{\pi_{\ast}(\psi,\theta)}{\pi(\psi,\theta)}d\psi d\theta.
\end{split}
\end{equation}
This gives the expression for the H-sensitivity.

Regarding the KL-sensitivity, it holds that 
\begin{align*}
\int & \pi(\eta|x)\log\left\{\frac{\pi(\eta|x)}{\pi_{\ast}(\eta|x)}\right\}d\eta\\
&=\int \pi(\eta|x)\log\left\{
\frac{\int f(x|\eta,\psi) \pi(\eta|\theta)\pi_{\ast}(\psi,\theta)d\psi d\theta / m_{\ast}(x)}
{\int f(x| \eta,\psi)\pi(\eta|\theta)\pi(\psi,\theta)d\psi d\theta / m(x) }
\right\} d\eta \\
&=
\int \pi(\eta|x)\log\left\{
\frac{\int f(x|\eta,\psi) \pi(\eta|\theta)\pi_{\ast}(\psi,\theta)d\psi d\theta }
{\int f(x| \eta,\psi)\pi(\eta|\theta)\pi(\psi,\theta)d\psi d\theta }
\right\} d\eta 
- 
\log\left\{\frac{m_{\ast}(x)}{m(x)}\right\}.
\end{align*}
Using (\ref{eq:ML-ratio4}) and (\ref{eq:ML-ratio2}) for the first and second terms, respectively, we obtain the result.

\bibliographystyle{chicago}
\bibliography{ref}

\end{document}